\documentclass[showpacs,preprintnumbers,amsmath,amssymb]{revtex4}
\usepackage{graphicx}% Include figure files
\usepackage{dcolumn}% Align table columns on decimal point
\usepackage{bm}% bold math
\usepackage{pstcol}
\usepackage{pst-coil}

\definecolor{light-blue}{rgb}{0.8,0.9,1}

\begin{document}

\preprint{UB-ZARM}

\title{Lorentz invariance violation and charge (non--)conservation: \\
A general theoretical frame for extensions of the Maxwell equations}

\author{Claus L\"ammerzahl}
\email{laemmerzahl@zarm.uni-bremen.de}
\affiliation{ZARM, University of Bremen, Am Fallturm, 28359 Bremen, Germany }

\author{Alfredo Mac\'{\i}as}
 \email{amac@xanum.uam.mx}
\affiliation{Universidad Autonoma Metropolitana Iztapalapa, Mexico}

\author{Holger M\"uller}
\email{holgerm@stanford.edu}
\affiliation{Physics Department, Stanford University, Stanford, CA 94305-4060, USA}

\date{\today}

\begin{abstract}
All quantum gravity approaches lead to small modifications in the
standard laws of physics which lead to violations of Lorentz
invariance. One particular example is the extended standard model
(SME). Here, a general phenomenological approach for extensions of
the Maxwell equations is presented which turns out to be more
general than the SME and which covers charge non--conservation
(CNC), too. The new Lorentz invariance violating terms cannot be
probed by optical experiments but need, instead, the exploration
of the electromagnetic field created by a point charge or a
magnetic dipole. Some scalar--tensor theories and higher
dimensional brane theories predict CNC in four dimensions and some
models violating Special Relativity have been shown to be
connected with CNC and its relation to the Einstein Equivalence
Principle has been discussed. Due to this upcoming interest, the
experimental status of electric charge conservation is reviewed.
Up to now there seem to exist no unique tests of charge
conservation. CNC is related to the precession of polarization, to
a modification of the $1/r$--Coulomb potential, and to a
time-dependence of the fine structure constant. This gives the
opportunity to describe a dedicated search for CNC.
\end{abstract}



\maketitle

\section{Introduction and Motivation}

The dynamics of the electromagnetic field is usually described by
the homogeneous and inhomogeneous Maxwell equations
\begin{equation}
\partial_{[\mu} F_{\nu\rho]} = 0 \, , \qquad \partial_\nu F^{\mu\nu}
= 4 \pi j^\mu \, , \label{ordinaryMaxwell}
\end{equation}
where $F_{\mu\nu} = \partial:_\mu A_\nu - \partial_\nu A_\mu$.
These equations can be based on charge conservation
\begin{equation}
\partial_\mu j^\mu = \dot\rho + \mbox{\boldmath$\nabla$} \cdot
\mbox{\boldmath$j$} = 0 \, . \label{CC}
\end{equation}
and on the conservation of the magnetic flux together with the
constitutive relation (see \cite{HehlObukhov03} for discussion of
formal structure of ordinary Maxwell equations). From
(\ref{ordinaryMaxwell}) it follows that light rays are
characterized by $\eta^{\mu\nu} k_\mu k_\nu = 0$ and propagate
along geodesics, and  that the charge is given by
\begin{equation}
Q = \int_\Sigma \j^0 d\Sigma = \int \rho d^3x\, ,
\end{equation}
where $\Sigma$ is some space--like hypersurface, is identified
with the total charge related to the 4--current density $j^\mu$.
(As usual, we assume that the charge and current density fall off
faster than $1/r^2$ for large space--like distances.) From
(\ref{CC}) we get $dQ/dt = 0$, where $t$ is the parameter
connected to the space--like hypersurfaces. Charge
non--conservation (CNC), on the other hand, would imply
\begin{equation}
\frac{d}{dt} Q \neq 0 \, .
\end{equation}
This is directly related to $\partial_\mu j^\mu \neq 0$.

The homogeneous Maxwell equations can be related to a conservation
of the magnetic flux \cite{HehlObukhov03} and serve as reason for
defining the Maxwell potential $A_\mu$, $F_{\mu\nu} = \partial_\mu
A_\nu - \partial_\nu F_\mu$. (For another relation to physical
phenomena, see below.) With the help of the homogeneous equations,
the inhomogeneous equations can be derived from a Lagrangian
${\cal L} = \frac{1}{16\pi} \eta^{\mu\rho} \eta^{\nu\sigma}
F_{\mu\nu} F_{\rho\sigma} + j^\mu A_\mu$ where $\eta^{\mu\nu}$ is
the Minkowski space--time metric. Charge conservation is a
consequence of the variational principle.

In the approach of Kosteleck\'{y} and coworkers
\cite{ColladayKostelecky98}, it has been suggested that
modifications of the effective Maxwell equations arising from
string theory are described by a replacement of
\begin{equation}
\eta^{\rho[\mu} \eta^{\nu]\sigma} \rightarrow \eta^{\rho[\mu} \eta^{\nu]\sigma}
+ k^{\mu\nu\rho\sigma} \label{replacement}
\end{equation}
in the ordinary Maxwell Lagrangian. This leads to a violation of
the Lorentz invariance of the physics of the SME which is related
to the coefficients encoded in the tensor $k^{\mu\nu\rho\sigma}$.
The question we want to address is whether this is the most
general extension of the ordinary Maxwell theory for the inclusion
of Lorentz violating terms. In the answer to this question we do
not start from a Lagrangian but, instead, use the field equations
only. Indeed, using this approach we are able to introduce further
Lorentz violating terms and, thereby, are able to include charge
non-conservation. The new Lorentz--violating extension cannot be
tested by optical experiments. We also think that, being such an
important part of present scheme of theoretical physics, charge
conservation needs also to be questioned and should be subject to
experimental tests in the same way as Lorentz invariance.

Charge conservation is a very important feature of ordinary Maxwell theory. It is
\begin{itemize}\itemsep=-2pt
\item basic for an interpretation of Maxwell--theory as $U(1)$
gauge theory, and \item it is necessary for the compatibility with
standard quantum theory in the sense that it is related to the
conservation of probability.
\end{itemize}
The more important a particular feature of physics is, the more
firmly this feature should be based on experimental facts.
Therefore, in this paper we are going to address the questions
\begin{itemize}\itemsep=-2pt
\item How good is charge conservation experimentally verified?
\item How to describe theoretically in a consistent way charge no-conservation?
\item Are there further consequences of charge non--conservation?
\item Do these consequences provide new tests of charge conservation?
\end{itemize}

It is also clear that, apart from its theoretical importance,
Lorentz invariance as well as charge conservation are central for
metrology, that is, for our current system of physical units and
the systems of fundamental constants. Any dependence of the
velocity of light from the velocity and from the orientation of
the laboratory as well as charge non--conservation will abolish
the uniqueness and universality of the definition of the second:
Each clock then will depend in its own way from the velocity and
orientation and, through the time--dependence of the fine
structure constant, from the used atomic transition. Related to
that is the definition of the meter and, of course, the definition
of all electrical units like the resistance and the voltage, both
based today on quantum phenomena, namely the von Klitzing effect
and the Josephson effect. Furthermore, as has been shown by Ni
\cite{Ni77} that any modification of the Maxwell equations (except
the one described by some axion field) will violate the validity
of the Universality of Free Fall, too, and is thus deeply
connected with the geometrization of the gravitational
interaction, and, consequently, important for the frame of General
Relativity. Since the ordinary Maxwell equations leads to Loretz
invariant effects and to charge conservation, any violation of
Lorentz invariance and CNC has necessarily to be described by a
modification of this set of Maxwell equations.

\section{Experimental Facts}

Since the experimental facts concerning the search for Lorentz
violation are well documented in every textbook we will restrict
ourselves to a few aspects of the experimental facts related to
charge conservation. There seem to be only three classes of
experiments related to charge conservation:

\begin{enumerate}
\item {\em Electron disappearing:} One aspect of charge
conservation is the spontaneous electron disappearing. This is
related to elementary particle decay processes like $e \rightarrow
\nu_e + \gamma$ or, more general, to $e \rightarrow$ any
particles. Decays of this kind have been searched for in processes
in high energy storage rings but nothing has been observed
\cite{Steinbergetal75,Aharonovetal95}. For the general process,
the probability for such a process has been estimated to be $2
\cdot 10^{-22}\;{\hbox{y}}^{-1}$ \cite{Steinbergetal75} and for
two specific processes the probability can be as less as $3 \cdot
10^{-26}\;{\hbox{y}}^{-1}$ \cite{Aharonovetal95}.

We note that even for a strict non--disappearing of electrons, the
charge of electron may vary in time and thus may give rise to CNC.
Therefore, while charge--conservation implies a non--disappearing
of electrons, electron non--disappearing does not imply charge
conservation.

\item {\it Equality of electron and proton charge:} Another aspect
of charge conservation is the equality of the absolute value of
the charge of all separable elementary particles like electrons
and protons (we let aside fractional charges of quarks because
these particles cannot be observed in isolated states). Tests of
the equality of $q_e$ and $q_p$ through the neutrality of atoms
\cite{DyllaKing73} yield very precise estimates. The reason for
that is that macroscopic numbers of atoms can be observed. The
experiment consists of the observation of sound waves in gas
induced by an externally applied time--dependent electric field
which should lead to a characteristic frequency of these sound
waves if there is an excess charge. The result is $|(q_e -
q_p)/q_e| \leq 10^{-19}$.

\item {\it Time--variation of $\alpha$:} The most direct test of
charge conservation is implied by searches for a time--dependence
of the fine structure constant $\alpha = q_e q_p/\hbar c$. Since
different hyperfine transitions depend in a different way on the
fine structure constant, a comparison of various transitions is
sensitive to a variation of $\alpha$. Recent comparisons of
different hyperfine transitions \cite{Marionetal03} lead to
$\left|\dot\alpha/\alpha\right| \leq 7.2 \cdot
10^{-16}\;{\hbox{y}}^{-1}$. This may be translated into an
estimate for charge conservation $\left|\dot q_e/q_e\right| \leq
3.6 \cdot 10^{-16}\;{\hbox{y}}^{-1}$, provided $\hbar$ and $c$ are
constant and $q_p = q_e$. However, this cannot be done within,
e.g., the frame of varying $c$ theories.

\end{enumerate}
Although there are individual experiments which have been used to
set limits on charge conservation, such as $e \rightarrow \nu_e +
\gamma$ \cite{Steinbergetal75,Aharonovetal95}, a general framework
is needed to allow the limits from different experiments to be
compared.

%\cite{Okun89}

\section{Models with violation of Lorentz invariance and CNC}

Models describing the violation of Lorentz invariance in the
context of the theory of electromagnetic fields have been
discussed since the early seventies. Ni
\cite{Ni73,Ni74,Ni77,Ni84,Ni84a,Ni87} considered a Lorentz
invariance violation in the Maxwell Lagrangian given by
(\ref{replacement}) and derived conditions on the coefficient
$k^{\mu\nu\rho\sigma}$ from the requirement that no birefringence
and no anisotropic speed of light should come out. Using this
approach he found a non--metric extension of Maxwell's theory
which is still compatible with the Weak Equivalence Principle
which, thus, constituted a counterexample to  Schiff's conjecture.
Lateron, Haugan and Kauffmann \cite{HauganKauffmann95} used the
same model in order to analyze astrophysical observations related
to birefringence. Recently, Kosteleck\'{y} and coworkers set up a
general scheme, the so--called Standard Model Extension (SME)
including the Maxwell as well as the fermion sector af particles
in order to describe violations of Lorentz invariance
\cite{ColladayKostelecky98}. For the Maxwell sector, again a
modification on the level of the Lagrangian has been used, and has
been confronted with astrophysical observations related to
birefringence and to laboratory experiments related to the
isotropy of the velocity of light
\cite{KosteleckyMewes01,KosteleckyMewes01} which, in the context
of a broad class of gravity theories, also lead to general time--
and position--dependent effects and violations of the Weak
Equivalence Principle \cite{Kostelecky04}. A simple model
including some mass vector for the photon and which already leads
to CNC has been introduced in \cite{LaemmerzahlHaugan01}.

Recently, some models which allow for a violation of charge
conservation have been discussed:  Within higher dimensional brane
theories it has been argued that charge may escape into other
dimensions \cite{DubovskyRubakovTinyakov00,DubovskyRubakov02} thus
leading to CNC in four--dimensional space--time. Also in
connection with variable--speed--of--light theories CNC may occur
\cite{LandauSisternaVucetich01}. A very important aspect of CNC is
its relation to the Einstein Equivalence Principle which is lying
at the basis of General Relativity
\cite{LandauSisternaVucetich01a}. This again emphasizes the fact
that the structure of Maxwell's equations is deeply connected with
the structure of space--time. CNC also necessarily appears if one
introduces phenomenologically a mass of the photon into the
Maxwell equations in a gauge--independent way
\cite{LaemmerzahlHaugan01}. This particular approach is also the
starting point for the considerations in this paper.

In this paper we do not want to proceed along the line of a
particular model but, instead, want to set up a general frame for
a theoretical description of CNC. The idea behind that is that any
violation of charge conservation must show up in a modification of
the Maxwell equations, and any modification of the Maxwell
equations should lead to a variety of effects which should be
accessible to various experimental tests. Such effects are, e.g.,
birefringence, dispersion, damping, anisotropy of the speed of
light, etc.

\section{How to measure charge? -- The definition of the electromagnetic field}

As a first step in the treatment of general Lorentz invariance
violations and CNC one first should clear how the electromagnetic
field and charges can be defined uniquely in an operational way.

\subsection{Charge in classical mechanics}

In the frame of classical mechanics the electric and magnetic
field as well as the charge is  measured using Lorentz' law
\cite{HehlObukhov03}
\begin{equation}
\mbox{\boldmath$F$} = q \, \mbox{\boldmath$E$}
+ \frac{q}{c} \mbox{\boldmath$v$} \times \mbox{\boldmath$B$} \, .
\end{equation}
However, if one allows for accelerating observers it is not
possible even to characterize uniquely what is a vanishing charge.
Furthermore, in this frame forces are probes of charge. If
charges, e.g., vary in time, then the force between charges varies
with time. This can be probed using e.g. springs. However, since
the physics of spring also heavily depends on the electromagnetic
field and the properties of the charges of the constituents, no
unique identification of charge conservation can be made.

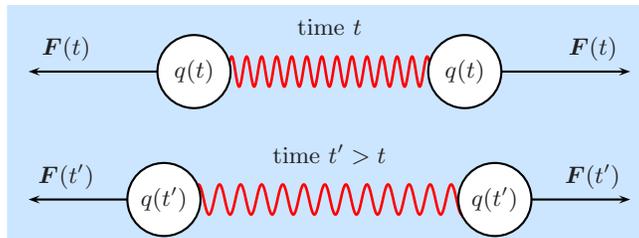
\begin{figure}[h]
\psset{unit=1cm}
\begin{center}
{
\begin{pspicture}(-4,-2)(4,1)
\psframe[linestyle=none,fillstyle=solid,fillcolor=light-blue](-4.3,-2)(4.3,1.2)
\rput(0,0.3){
\rput(0,0.6){time $t$}
\psline{->}(1.8,0)(4,0)
\psline{->}(-1.8,0)(-4,0)
\pscoil[coilaspect=0,coilheight=0.5,coilwidth=0.4,linewidth=1pt,linecolor=red](-1.9,0)(1.85,0)
\pscircle[fillstyle=solid,fillcolor=white](-1.8,0){0.5}
\pscircle[fillstyle=solid,fillcolor=white](1.8,0){0.5}
\rput(-1.8,0){$q(t)$}
\rput(1.8,0){$q(t)$}
\rput(3.5,0.3){$\mbox{\boldmath$F$}(t)$}
\rput(-3.5,0.3){$\mbox{\boldmath$F$}(t)$}
}
\rput(0,-1.4){
\rput(0,0.6){time $t^\prime > t$}
\psline{->}(2,0)(4,0)
\psline{->}(-2,0)(-4,0)
\pscoil[coilaspect=0,coilheight=0.7,coilwidth=0.4,linewidth=1pt,linecolor=red](-2.3,0)(2.25,0)
\pscircle[fillstyle=solid,fillcolor=white](-2.2,0){0.5}
\pscircle[fillstyle=solid,fillcolor=white](2.2,0){0.5}
\rput(-2.2,0){$q(t^\prime)$}
\rput(2.2,0){$q(t^\prime)$}
\rput(3.5,0.3){$\mbox{\boldmath$F$}(t^\prime)$}
\rput(-3.5,0.3){$\mbox{\boldmath$F$}(t^\prime)$}
}
%\showgrid
\end{pspicture}
}
\end{center}
\caption{Forces between increasing charges can be probed with a
spring.}
\end{figure}

\subsection{Charges in quantum mechanics}

A more fundamental approach is the physics of the hydrogen atom,
which can be used to probe the electromagnetic interaction between
charged particles on the quantum level. Here, charge comes in via
\begin{equation}
E = - \hbox{Ry} \frac{1}{n^2}\, , \quad \hbox{Ry} = \frac{\mu q_e^2 q_p^2}{\hbar^2}\, ,
\quad \mu = \frac{m_e  m_p}{m_e + m_p} \, ,
\end{equation}
where $q_e$, $q_p$, $m_e$, and $m_p$ are the charges and masses of
the electron and the proton, respectively. For other levels, like
hyperfine levels, the expressions are more complicated but still
depend on the same set of constants.

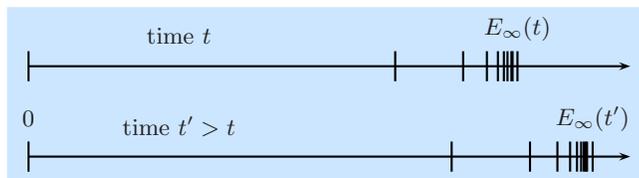
\begin{figure}[h]
\psset{unit=1cm}
\begin{center}
{
\begin{pspicture}(-4,-1)(4,1)
\psframe[linestyle=none,fillstyle=solid,fillcolor=light-blue](-4.3,-1.2)(4.3,1.2)
\rput(0,0.2){
\rput(-2,0.6){time $t$}
\psline{->}(-4,0.2)(4,0.2)
\psline(-4,0)(-4,0.4)
\psline(0.875,0)(0.875,0.4)
\psline(1.77778,0)(1.77778,0.4)
\psline(2.09375,0)(2.09375,0.4)
\psline(2.24,0)(2.24,0.4)
\psline(2.31944,0)(2.31944,0.4)
\psline(2.36735,0)(2.36735,0.4)
\psline(2.41975,0)(2.41975,0.4)
\psline(2.435,0)(2.435,0.4)
\psline(2.5,0)(2.5,0.4)
\rput(2.5,0.7){$E_\infty(t)$}
}
\rput(0,-1){
\rput(-2,0.6){time $t^\prime > t$}
\psline{->}(-4,0.2)(4,0.2)
\psline(-4,0)(-4,0.4)
\psline(1.625,0)(1.625,0.4)
\psline(2.66667,0)(2.66667,0.4)
\psline(3.03125,0)(3.03125,0.4)
\psline(3.2,0)(3.2,0.4)
\psline(3.29167,0)(3.29167,0.4)
\psline(3.34694,0)(3.34694,0.4)
\psline(3.38281,0)(3.38281,0.4)
\psline(3.40741,0)(3.40741,0.4)
\psline(3.425,0)(3.425,0.4)
\psline(3.5,0)(3.5,0.4)
\rput(3.5,0.7){$E_\infty(t^\prime)$}
}
 \rput(-4,-0.3){0}
%\showgrid
\end{pspicture}
}
\end{center}
\caption{Atomic energy levels (here the Balmer series, for
example) are sensitive to increasing charges.}
\end{figure}

The important point in both cases is that one needs a reference
''system'', either a spring (depends on electric properties, too),
or quantum mechanics. Since also the spring basically is
determined by quantum properties, both reference systems depend,
beside the charge, on other parameters like $\hbar$, $c$, $m_e$,
$m_p$. Therefore, these methods are not applicable for a unique
and independent definition of charge.

We may ask the following question: If we start from the ordinary
Schr\"odinger equation minimally coupled to the electromagnetic
field, what parameter is effectively probed by experiments
described by the Schr\"odinger equation? If we rewrite the
Schr\"odinger equation in the form
\begin{equation}
i \frac{\partial}{\partial t} \psi = - \frac{1}{2 m/\hbar} \left(\nabla
- \frac{i q_e}{\hbar c} A\right)^2 - \frac{m}{\hbar} U \psi
+ \frac{q_e}{\hbar} \phi \psi \, , \label{Schroedinger}
\end{equation}
then it is clear that there are two effective parameters,
$m/\hbar$ and $q_e/\hbar$ where $m/\hbar$ can be determined with
high precision with neutron or atom interferometry. In order to
determine the charge or the mass of the quantum particle, one has
to know $\hbar$. This quantity today can be measured best using
the Watt balance \cite{Williamsetal98}. However, this is based on
the primary standard kilogram, on the measurement of the
gravitational acceleration as well as on the units of electrical
resistance and voltage defined through the von--Klitzing effect
and the Josephson effect which, in turn, again depends on the
charge. Since the thorough analysis of all the dependence between
the various definitions is beyond this paper, we use $m/\hbar$ and
$q_e/\hbar$ as (effective) mass and charge.

Inspired by the structure of the Schr\"odinger equation we may
proceed with an operational definition of charge and of the
electromagnetic field using the Aharonov--Bohm effect.

\subsection{Definition of charge and of the electromagnetic field}

In order to develop a general frame for describing in a consistent
way the dynamics of the electromagnetic field and, thus,
violations of Lorentz invariance and CNC, we first define the
electromagnetic field strength $F_{\mu\nu}$ and the electric
charge. This will be done through the Aha\-ronov--Bohm effect for
charged quantum particles.

If we perform an interference experiment for charged particles in
an electromagnetic field, then we make the basic experience that
the phase shift depends in a linear way on the area enclosed by
the particles' path. For doing this observation we do not need any
description of the electromagnetic field. We just need to have
some apparatus which produces some electromagnetic field which we
also should be able to vary. This basic experience can be encoded
in the general formula ($\mu, \nu, \ldots$ are indices ranging
from $0, \ldots, 3$)
\begin{equation}
\Delta\phi = \frac{q}{\hbar} \int F_{\mu\nu} d\sigma^{\mu\nu}\, , \label{ABphase}
\end{equation}
where, $\sigma^{\mu\nu}$ is the area element which is an
antisymmetric tensor. For given area and measured phase, this
equation defines the electromagnetic field $F_{\mu\nu}$. This
phase shift, of course, coincides with the phase shift for charged
particle when using (\ref{Schroedinger}). However, here we use
(\ref{ABphase}) independent from (\ref{Schroedinger}).

By observing this phase shift for different particles in the same
experimental situation, we recognize that the linear dependence of
the phase shift on the area which differs only by a
proportionality factor $q/\hbar$. This proportionality factor we
call the {\it charge} of the particle. This charge can indeed be
defined uniquely. The reason for that is that in nature there are
neutral particles which, by means of our approach, can be
identified uniquely: Neutral particles are defined as those
particles whose interference pattern does not vary in the case of
a varying electromagnetic field (though the Aharonov--Bohm effect
is used for the definition of the electromagnetic field, this is
no logical circle because for the identification of neutral
particles we do not need the quantitative definition of the
electromagnetic field, we just need to vary it in some way). The
fact that here we are able to uniquely identify neutral particles
while this is not possible by using the Lorentz force equation
only shows that the Aharonov-Bohm effect is of superior
importance.

After having identified neutral particles, we are in the position
to identify particles with different charges $q$ by merely
comparing the phases for different particles travelling along the
same geometrical path: $q_1/q_2 = \Delta\phi_1/\Delta\phi_2$. By
consideration of all kinds of charged elementary particles, we can
identify that one with the smallest charge, and it is also a basic
experience, again from the Aharonov--Bohm effect, that all
particles come with charges which is a multiple of the elementary
charge $e$, that is, particles have a discrete spectrum of charges
only. However, the fact that all particles have charges $q/\hbar =
n e/\hbar$, $n \in \mathbb{N}$, does not mean that charge is
conserved. Indeed, it is still possible that the elementary charge
in $e/\hbar$ varies with time. In this case, all the observations
connected with charged particle interferometry and derived notions
are still consistent. However, as we shall see later, a varying
charge is not consistent with the ordinary Maxwell equations.

As a first result, the uniqueness of the phase shift
(\ref{ABphase}) requires (for simplicity, we exclude any
non--trivial topology of the underlying space--time)
\begin{equation}
\partial_{[\mu} F_{\rho\sigma]} = 0 \, . \label{homMaxwell}
\end{equation}
what establishes the homogenous Maxwell equations.

\section{A general frame for the dynamics of electromagnetic fields}

\subsection{The general ansatz}

The homogeneous equations (\ref{homMaxwell}) are dynamically
incomplete because they provide only three dynamical equations for
the six components encoded in $F_{\mu\nu}$. Therefore
(\ref{homMaxwell}) has to be completed by another set containing
three further dynamical equations. In a pragmatic approach, we
start from the ordinary inhomogenous Maxwell equations which are
experimentally verified to a very high degree of accuracy
\cite{Laemmerzahl01}. Therefore, if in nature there is some
violation of Lorentz invariance or CNC, then this must be a tiny
correction to the ordinary equations. The most general
modification of the inhomogeneous Maxwell equations for field
strength $F$ which still is linear in $F$ and first order in
derivative is given by
\begin{equation}
(\eta^{\mu\rho} \eta^{\nu\sigma} + \chi^{\mu\nu\rho\sigma})
\partial_\nu F_{\rho\sigma} + \chi^{\mu\rho\sigma} F_{\rho\sigma}
= 4 \pi j^\mu \, . \label{GME}
\end{equation}
Here, $\eta^{\mu\rho} \eta^{\nu\sigma} + \chi^{\mu\nu\rho\sigma}$
plays the role of a generalized constitutive or material tensor
and $\chi^{\mu\rho\sigma}$ possess properties of an anisotropic
mass of the photon. $\eta^{\mu\nu}$ is the Minkowksi metric with
signature $(+,-,-,-)$. We assume the $\chi^{\mu\nu\rho\sigma}$ and
$\chi^{\mu\rho\sigma}$ to be constant. It is no problem to
generalize this approach to the case of a Riemannian metric
$\eta^{\mu\nu}$ instead of the Minkowski metric and to consider
position dependent tensors $\chi^{\mu\nu\rho\sigma}$ and
$\chi^{\mu\rho\sigma}$. We restrict ourselves to this more simple
case since, as far as Lorentz invariance is concerned, all
conclusions will be the same. --- The charge underlying the
4--current on the right hand side is identified with the charge
which has been defined using the Aharonov--Bohm effect. That means
that we identify active and passive electromagnetic charges.

Eq.(\ref{GME}) is our model describing violations of Lorentz
invariance and CNC. Before discussing Lorentz non--invariance and
CNC, we make a few general comments:
\begin{itemize}
\item The generalized constitutive tensor possess the symmetry
\begin{equation}
\chi^{\mu\nu\rho\sigma} = \chi^{\mu\nu[\rho\sigma]}
\end{equation}
and, thus, possesses 96 components. Below we impose two more
conditions on this constitutive tensor: first
$\chi^{\mu[\nu\rho\sigma]} = 0$ which eliminates the degrees of
freedom which will drop out from the inhomogenous Maxwell
equations (\ref{GME}) due to the homogeneous ones
(\ref{homMaxwell}). Furthermore, the uniqueness of the
Cauchy--problem requires $\chi^{(\alpha\beta\mu)\nu} = 0$. This
reduces the number of independent coefficients to 35. One of these
components can be absorbed by a redefinition of the charge. This
generalized constitutive tensor implies birefringence, an
anisotropic velocity of light, an anisotropic Coulomb potential
and a modified magnetic field. \item In equations which use field
strengths only it is not possible to assign a scalar mass to the
photon. Masses have to be tensor valued. This is in contrast to
the Proca equation which is used as model for describing a scalar
photon mass. \item The mass tensor possesses the symmetry
$\chi^{\mu\rho\sigma} = \chi^{\mu[\rho\sigma]}$ and, thus, 24
independent components. It leads an anisotropy and velocity
depencence of the velocity of light, to birefringence, dispersion,
a damping of the intensity of radiation, and to an  anisotropy and
a Yukawa modification of the Coulomb potential. \item Our model,
in general, cannot be derived from a Lagrangian. In fact, in order
to get the second term, the corresponding term in the Lagrangian
has to be linear in $\chi^{\mu\rho\sigma}$ and quadratic in the
potential $A_\mu$. The only combination is $\chi^{\mu\rho\sigma}
A_\mu F_{\rho\sigma}$. This term is gauge invariant only, if
$\chi^{\mu\rho\sigma} \partial_\mu F_{\rho\sigma}$ vanishes. This
is the case only for $\chi^{\mu\rho\sigma} =
\chi^{[\mu\rho\sigma]}$.
\end{itemize}

\subsection{Decomposition of the constitutive tensor}

First, we introduce the decomposition of the general constitutive
tensor into irreducible parts (see Appendix for the definitions)
\begin{eqnarray}
\chi^{\alpha\beta\mu\nu} & = & {}^{(1)}W^{\alpha\beta\mu\nu}
+ \frac{1}{2} \epsilon^{\mu\nu}{}_{\rho\sigma} \eta^{\rho[\alpha}
\Psi^{\beta]\sigma} + \frac{1}{12} X \epsilon^{\alpha\beta\mu\nu}
- \eta^{\mu[\alpha} \Phi^{\beta]\nu} + \eta^{\nu[\alpha} \Phi^{\beta]\mu} \nonumber\\
& & + \eta^{\mu[\alpha} W_{\rm a}^{\beta]\nu}
- \eta^{\nu[\alpha} W_{\rm a}^{\beta]\mu}
- \frac{1}{6} W \eta^{\alpha [\mu} \eta^{|\beta|\nu]}
+ {}^{(1)}Z^{\alpha\beta\mu\nu}
+ \frac{1}{2} \epsilon^{\mu\nu}{}_{\rho\sigma} \eta^{\rho(\alpha}
\Upsilon^{\beta)\sigma} \nonumber\\
& &  + \frac{1}{3} \left(2 \eta^{\mu(\alpha} \Delta^{\beta)\nu}
- 2 \eta^{\nu(\alpha} \Delta^{\beta)\mu} - \eta^{\alpha\beta} \Delta^{\mu\nu}\right)
+ \frac{1}{4} \eta^{\alpha\beta} Z^{\mu\nu} \nonumber\\
& & + \frac{1}{2} \left(\eta^{\mu(\alpha} \Xi^{\beta)\nu}
- \eta^{\nu(\alpha} \Xi^{\beta)\mu}\right)  \label{ConstGeneral}
\end{eqnarray}
The generalized constitutive tensor is more general than the
ordinary constitutive tensor which possesses the symmetries of the
Riemann tensor, namely
\begin{equation}
k^{\mu\nu\rho\sigma} = k^{[\mu\nu]\rho\sigma} = k^{\mu\nu[\rho\sigma]}
= k^{\rho\sigma\mu\nu} \, , \qquad k^{\mu[\nu\rho\sigma]} = 0 \label{SymmOrdConst}
\end{equation}
and, thus, consists of 20 components. Within a Lagrange--ansatz
${\cal L} = \frac{1}{16\pi} \left(\eta^{\mu\nu} \eta^{\nu\sigma} +
k^{\mu\nu\rho\sigma}\right) F_{\mu\nu} F_{\rho\sigma}$ the totally
antisymmetric part results in a divergence and the scalar part can
be absorbed into a redefinition of the coupling to matter, so that
$k^{\mu\nu\rho\sigma}$ effectively possesses 19 components only.
The implications of $k^{\mu\nu\rho\sigma}$ have been described by
Ni \cite{Ni73,Ni74,Ni77,Ni84}, Haugan and Kauffmann
\cite{HauganKauffmann95} and recently by Kosteleck\'{y}, Mewes and
others \cite{KosteleckyMewes01,KosteleckyMewes02,Muelleretal03}.

The ''mass'' term can be decomposed into three irreducible parts,
the totally antisymmetric part, a trace and the rest
\begin{equation}
\chi^{\alpha\mu\nu} = {}^{(1)}\chi^{\alpha\mu\nu}
+ \epsilon^{\alpha\beta\mu\nu} a_\beta + \eta^{\alpha[\mu} t^{\nu]}  \label{irredmass}
\end{equation}
where $a_\beta = \frac{1}{6} \epsilon_{\alpha\beta\mu\nu}
\chi^{\alpha\mu\nu}$, $t^\nu = \frac{2}{3} \eta_{\alpha\mu}
\chi^{\alpha\mu\nu}$, and ${}^{(1)}\chi^{\alpha\mu\nu} =
\chi^{\alpha\mu\nu} - \epsilon^{\alpha\beta\mu\nu} a_\beta -
\eta^{\alpha[\mu} t^{\nu]}$. Both vectors $a_\mu$ and $t^\mu$
possess four and ${}^{(1)}\chi^{\alpha\mu\nu}$ possesses 16
components.

\subsection{Unique time evolution}

A $3+1$--decomposition of the generalized Maxwell equations (\ref{GME}) gives
\begin{eqnarray}
4 \pi \rho & = & \mbox{\boldmath$\nabla$} \cdot \mbox{\boldmath$E$}
+ \chi^{000i} \dot E_i + \chi^{00ij} \dot B_{ij} + \chi^{0i\rho\sigma}
\partial_i F_{\rho\sigma} + \chi^{0\rho\sigma} F_{\rho\sigma}\label{Inhom1} \\
4 \pi j^i & = & - (\mbox{\boldmath$\nabla$} \times \mbox{\boldmath$B$})^i
+ \dot E_i + \chi^{i00j} \dot E_j + \chi^{i0jk} \dot B_{jk}
+ \chi^{ij\rho\sigma} \partial_j F_{\rho\sigma}
+ \chi^{i\rho\sigma} F_{\rho\sigma} \, , \label{Inhom2}
\end{eqnarray}
where $E_i = F_{0i}$, $B_{ij} = F_{ij}$ and $B_i = \frac{1}{2}
\epsilon_{ijk} B_{jk}$. The time derivative on
$\mbox{\boldmath$B$}$ can be replaced by a spatial derivative of
the electric field with the help of the homogenous equations. The
important point is, that, owing to the term $\chi^{000i}$, both
equations are dynamical equations for the electric field. In order
that the generalized Maxwell equation leads to a consistent
dynamical description, we have to require the vanishing of the
coefficient $\chi^{000i}$ (since we want to recover the ordinary
Maxwell equations for vanishing $\chi^{\mu\nu\rho\sigma}$ we
cannot choose $\chi^{i00j} = - \delta^{ij}$ though this, in
principle, would also solve the problem). Since this should be
true for any chosen frame of reference, we have to require
\begin{equation}
\chi^{(\mu\nu\rho)\sigma} = 0 \, .  \label{ConsistencyCondition}
\end{equation}

This consistency requirement can also be read off directly from
the following observation: Solving the second equation for $\dot
E_i$ and inserting this into the first equation (only keeping
terms first order in the $\chi$'s), gives the condition
\begin{equation}
4 \pi \rho = \mbox{\boldmath$\nabla$} \cdot \mbox{\boldmath$E$}
+ \chi^{000i} \left(j^i - (\mbox{\boldmath$\nabla$}
\times \mbox{\boldmath$B$})^i\right) + \chi^{00ij} \partial_i E_j
+ \chi^{0i\rho\sigma} \partial_i F_{\rho\sigma}  + \chi^{0\rho\sigma} F_{\rho\sigma} \, .
\end{equation}
For given charge density $\rho$ and current $\mbox{\boldmath$j$}$
and given initial conditions on $\mbox{\boldmath$E$}$ and
$\mbox{\boldmath$B$}$, this represents an additional condition on
the sources implying that one is not free to choose arbitrary
sources. Since this is non--physical, we have to require
(\ref{ConsistencyCondition}).

The requirement (\ref{ConsistencyCondition}) imposes conditions on
the general constitutive tensor. These conditions are readily
shown to be
\begin{equation}
\chi^{(\alpha\beta\mu)\nu} = 0 \qquad \Rightarrow \qquad
{}^{(1)}Z_{(\alpha\beta\mu)\nu} = 0 \, , \quad \Xi_{\mu\nu} = 0 \, , \quad \Delta_{\mu\nu} = - \frac{3}{4} Z_{\mu\nu} \, .
\end{equation}
The decomposition of the constitutive tensor then simplifies to be
\begin{eqnarray}
\chi^{\alpha\beta\mu\nu} & = &  {}^{(1)}W^{\alpha\beta\mu\nu}
+ \frac{1}{2} \epsilon^{\mu\nu}{}_{\rho\sigma} \eta^{\rho[\alpha} \Psi^{\beta]\sigma}
+ \frac{1}{12} X \epsilon^{\alpha\beta\mu\nu} - \eta^{\mu[\alpha} \Phi^{\beta]\nu}
+ \eta^{\nu[\alpha} \Phi^{\beta]\mu} \nonumber\\
& & + \eta^{\mu[\alpha} W_{\rm a}^{\beta]\nu} - \eta^{\nu[\alpha} W_{\rm a}^{\beta]\mu}
- \frac{1}{6} W \eta^{\alpha [\mu} \eta^{|\beta|\nu]} \nonumber\\
& & + \frac{1}{2} \epsilon^{\mu\nu}{}_{\rho\sigma} \eta^{\rho(\alpha}
\Upsilon^{\beta)\sigma} - \frac{1}{2} \left(\eta^{\mu(\alpha} Z^{\beta)\nu}
- \eta^{\nu(\alpha} Z^{\beta)\mu}\right) + \frac{1}{2} \eta^{\alpha\beta} Z^{\mu\nu}
\label{ctidCp}
\end{eqnarray}
and the corresponding Maxwell equations then read
\begin{eqnarray}
\chi^{\alpha\beta\mu\nu} \partial_\beta F_{\mu\nu} & = & \partial_\nu F^{\mu\nu}
+ {}^{(1)}W^{\alpha\beta\mu\nu} \partial_\beta F_{\mu\nu}
- 2 \eta^{\mu[\alpha} \Phi^{\beta]\nu} \partial_\beta F_{\mu\nu}  \nonumber\\
& & + \left(\eta^{\alpha\mu} W^{[\beta\nu]} - \eta^{\beta\mu} W^{[\alpha\nu]}\right)
\partial_\beta F_{\mu\nu} - \frac{1}{6} W \eta^{\alpha\mu} \eta^{\beta\nu}
\partial_\beta F_{\mu\nu} \nonumber\\
& & + \frac{1}{2} \epsilon^{\mu\nu}{}_{\rho\sigma} \left(\eta^{\rho(\alpha}
\Upsilon^{\beta)\sigma} + \eta^{\rho[\alpha} \Psi^{\beta]\sigma}\right)
\partial_\beta F_{\mu\nu} - \eta^{\mu(\alpha} Z^{\beta)\nu} \partial_\beta F_{\mu\nu}
+ \frac{1}{2} \eta^{\alpha\beta} Z^{\mu\nu} \partial_\beta F_{\mu\nu} \, .
\end{eqnarray}
It is easy to see that neither the $\Upsilon$-- nor the $Z$--terms
contribute to a time derivative of the electric field.

\subsection{Charge (non--)conservation}

Within our model (\ref{GME}) the divergence of the 4--current in
general is not assumed to vanish
\begin{equation}
4 \pi \partial_\mu j^\mu = \chi^{\mu\nu\rho\sigma} \partial_\mu
\partial_\nu F_{\rho\sigma} + \chi^{\mu\rho\sigma} \partial_\mu F_{\rho\sigma} \, .
\end{equation}
However, it is amazing that already at this stage, that is using
(\ref{ConsistencyCondition}) only, the principal part of the
Maxwell equations respect charge conservation: Employing the above
decomposition we get for the divergence of the anomalous part of
the principal part
\begin{eqnarray}
\partial_\alpha (\chi^{\alpha\beta\mu\nu} \partial_\beta F_{\mu\nu}) & = & \frac{3}{8}
\epsilon^{\mu\nu\alpha}{}_{\sigma} \Psi^{\beta\sigma} \partial_\alpha \partial_\beta
F_{\mu\nu} - \frac{1}{2} \eta^{\alpha\mu} Z^{\nu\beta} \partial_\alpha \partial_\beta
F_{\mu\nu} \nonumber\\
& & + \frac{3}{2} \eta^{\beta\mu} Z^{\nu\alpha} \partial_\alpha \partial_\beta
F_{\mu\nu} + \frac{1}{2} \eta^{\alpha\beta} Z^{\mu\nu} \partial_\alpha \partial_\beta
F_{\mu\nu} + \chi^{\mu\rho\sigma} \partial_\mu F_{\rho\sigma} \, ,
\end{eqnarray}
where we showed those terms only which do not vanish trivially due
to the antisymmetry in $\alpha$ and $\beta$. Here, the $\Psi$
vanishes because of its total antisymmetry in $\alpha, \mu, \nu$.
The $Z$--terms can be treated as follows
\begin{eqnarray}
Z-\hbox{terms} & = & \eta^{\alpha\mu} Z^{\nu\beta} \partial_\alpha \partial_\beta
F_{\mu\nu} + \frac{1}{2} \eta^{\alpha\beta} Z^{\mu\nu} \partial_\alpha \partial_\beta
F_{\mu\nu} \nonumber\\
& = & \eta^{\alpha\beta} Z^{\mu\nu} \partial_\alpha \left(- 2 \partial_\mu F_{\beta\mu}
+ \partial_\beta F_{\mu\nu}\right) \nonumber\\
& = & 0
\end{eqnarray}
due to the homogenous Maxwell equations. Therefore, even for the
very general ansatz (\ref{GME}) with (\ref{ConstGeneral}) the {\it
dynamical consistency automatically ensures charge conservation
for the principal part of the generalized Maxwell equations}.
Thus, charge non--conservation can arise from the
$\chi^{\alpha\mu\nu}$ only
\begin{equation}
4 \pi \partial_\alpha j^\alpha = \chi^{\alpha\mu\nu} \partial_\alpha F_{\mu\nu}
= {}^{(1)}\chi^{\alpha\mu\nu} \partial_\alpha F_{\mu\nu} + \eta^{\alpha\mu}
t^\nu \partial_\alpha F_{\mu\nu} \, .
\end{equation}
The totally antisymmetric part of $\chi^{\alpha\mu\nu} =
\epsilon^{\alpha\beta\mu\nu} a_\beta$ still is compatible to
charge conservation\footnote{In a Lagrangian formulation such a
term comes from the totally antisymmetric part of the constitutive
tensor which then, however, has to be position dependent: Indeed,
taking as part of the Lagrangian $\theta F_{\mu\nu}
\epsilon^{\mu\nu\rho\sigma} F_{\rho\sigma}$, then this leads to a
term $\partial_\nu \theta \epsilon^{\mu\nu\rho\sigma}
F_{\rho\sigma}$ in the inhomogenous Maxwell equations. This is the
axion as introduced by Ni \cite{Ni73} establishing a
counterexample to Schiff's conjecture. On the level of the field
equations the axion is not part of the constitutive tensor but,
instead, part of the ''mass'' tensor.}. The quantity $a_\beta$ is
the axion introduced first by Ni \cite{Ni73}. All other parts of
$\chi^{\alpha\mu\nu}$ lead to CNC.

The change of the total charge can be expressed in terms of the
tensor $\chi^{\mu\rho\sigma}$,
\begin{eqnarray}
\frac{dQ}{dt} & = & \lim_{\delta t \rightarrow 0} \frac{1}{\delta t}
\left(\int_{t + \delta t} \rho d^3 x - \int_t \rho d^3 x\right) \nonumber\\
& = & \int \partial_\mu j^\mu d^3 x \nonumber\\
& = & \int \left({}^{(1)}\chi^{\mu\rho\sigma} \partial_\mu F_{\rho\sigma}
+ \eta^{\mu\rho} t^\sigma  \partial_\mu F_{\rho\sigma}\right) d^3 x \, . \label{cnc1}
\end{eqnarray}
The variation of the charge depends on the actual solution
$F_{\mu\nu}$ of the generalized Maxwell equations. In the case
${}^{(1)}\chi^{\mu\rho\sigma} = 0$ this simplifies considerably
\begin{equation}
\frac{dQ}{dt} = t_0 Q + \int \mbox{\boldmath$t$} \cdot \mbox{\boldmath$j$} d^3 x \, ,
\end{equation}
where we assumed that the charge density falls off fast enough at
spatial infinity. If there are no currents present, then the
change of the charge can be given directly in terms of the charge
solely, $\frac{dQ}{dt} = t_0 Q$ \cite{LaemmerzahlHaugan01}.

It is obvious that already in this simple case there is no static
solution for the electric field of a point charge. That means, in
particular, that we do not have energy conservation in the sense
of an invariance of the solution with respect to time
translations. One then might think of testing CNC parameters by
analyzing energy conservation in high energy experiments. This
will not lead to good results since in scattering processes the
time scale is too short for probing the time dependence of the
energy or the charge. For our purposes, long term experiments are
needed as, e.g., searches for a time--dependence of the fine
structure constant.

\subsection{Use of the homogeneous Maxwell equations}

Because of (\ref{homMaxwell}) the corresponding part in the
inhomogenous Maxwell equations, $\chi^{\alpha[\beta\rho\sigma]}
\partial_\beta F_{\rho\sigma}$, will vanish identically and drops
out of the inhomogeneous Maxwell equations. The corresponding
parts of the irreducible decomposition play no role and can be
assumed to vanish, $\chi^{\alpha[\beta\mu\nu]} = 0$ or,
equivalently, $\epsilon_{\delta\beta\mu\nu}
\chi^{\alpha[\beta\mu\nu]} = 0$. From the irreducible
decomposition (\ref{ctidCp}) we get
\begin{equation}
\epsilon_{\delta\beta\mu\nu} \chi^{\alpha[\beta\mu\nu]}
= - \frac{1}{2} X \delta_\delta^\alpha - \epsilon^\alpha{}_{\delta\mu\nu}
(Z - W_{\rm a})^{\mu\nu} + (2 \Upsilon - \Psi)^{\alpha}{}_{\delta} \, .
\end{equation}
That means
\begin{equation}
\chi^{\alpha[\beta\mu\nu]} = 0 \qquad \Rightarrow \qquad X = 0 \, , \quad Z^{\mu\nu}
= W_{\rm a}^{\mu\nu}\, , \quad \Upsilon^{\mu\nu} = \frac{1}{2} \Psi^{\mu\nu} \, .
\end{equation}
With this result the constitutive tensor reduces to
\begin{eqnarray}
\chi^{\alpha\beta\mu\nu} & = &  {}^{(1)}W^{\alpha\beta\mu\nu}
+ \frac{1}{8} \epsilon^{\mu\nu}{}_{\rho\sigma} \left(3 \eta^{\rho\alpha}
\Psi^{\beta\sigma} - \eta^{\rho\beta} \Psi^{\alpha\sigma}\right)
- \eta^{\mu[\alpha} \Phi^{\beta]\nu} + \eta^{\nu[\alpha} \Phi^{\beta]\mu} \nonumber\\
& & - \frac{1}{2} \eta^{\alpha[\mu} Z^{\nu]\beta}
+ \frac{3}{2} \eta^{\beta[\mu} Z^{\nu]\alpha} + \frac{1}{2}
\eta^{\alpha\beta} Z^{\mu\nu} - \frac{1}{6} W \eta^{\alpha [\mu} \eta^{|\beta|\nu]}
\label{ctcphm}
\end{eqnarray}
Now the effective Maxwell equations are
\begin{eqnarray}
4 \pi j^\alpha & = & \partial_\nu F^{\mu\nu} + \chi^{\alpha\beta\mu\nu}
\partial_\beta F_{\mu\nu} \nonumber\\
& = & \partial_\nu F^{\mu\nu} + {}^{(1)}W^{\alpha\beta\mu\nu}
\partial_\beta F_{\mu\nu} + \frac{3}{8} \epsilon^{\mu\nu\alpha}{}_{\sigma}
\Psi^{\beta\sigma} \partial_\beta F_{\mu\nu} - 2 \eta^{\mu[\alpha}
\Phi^{\beta]\nu} \partial_\beta F_{\mu\nu} \nonumber\\
& & - \frac{1}{2} \eta^{\alpha\mu} Z^{\nu\beta} \partial_\beta F_{\mu\nu}
+ \frac{3}{2} \eta^{\beta\mu} Z^{\nu\alpha} \partial_\beta F_{\mu\nu}
+ \frac{1}{2} \eta^{\alpha\beta} Z^{\mu\nu} \partial_\beta F_{\mu\nu}
- \frac{1}{6} W \eta^{\alpha \mu} \eta^{\beta\nu} \partial_\beta F_{\mu\nu} \, . \label{GME2}
\end{eqnarray}

Since these Maxwell equations are more general than those in the
extended standard model we now have to look anew for ways how to
confront these equations with the experiment in order to give
unique experimental criteria for the Lorentz invariance of the
theory. It has been shown by Ni \cite{Ni77} and Kosteleck\'{y} and
Mewes \cite{KosteleckyMewes01,KosteleckyMewes02} that for the SME
the requirement of vanishing birefringence and isotropic speed of
light leads to a Lorentz invariant  theory. The question now is
whether this remains true for the present framework. Compared with
the Lagrangian based approach by Ni, Haugan, Kosteleck\'{y} and
others, we are more general by the terms $\Psi^{\mu\nu}$ and
$Z^{\mu\nu}$.

\section{Propagation of light}

Now we want to analyze some physical consequences of the
generalized Maxwell equations (\ref{GME2}). A first step is to
determine the wave equation for the electromagnetic field, to
calculate the dispersion relation and to discuss consequences like
birefringence and anisotropic speed of light. Also the propagation
of the polarization is sensitive to effects connected with the
violation of Lorentz invariance. We will see, that not all Lorentz
invariance violating terms are accessible by radiation effects,
that is, by optical experiments: Therefore, there is a need to
discuss also the static electromagnetic fields of point charges
and magnetic moments (Sec.\ref{PointCharges}).

\subsection{The wave equation}

We derive the wave equation in vacuum in the usual way by
differentiating Eq.(\ref{GME2}) and substituting the time
derivative of the magnetic field using the homogeneous equation
\begin{equation}
0 = \ddot E_i - \Delta E_i + (\mbox{\boldmath$\nabla$} (\mbox{\boldmath$\nabla$}
\cdot \mbox{\boldmath$E$}))^i + \chi^{i\mu\nu j} \partial_\mu \partial_\nu E_j
+ 2 \chi^{i0j} \dot E_i + \chi^{ikj} \partial_k E_j \label{ModWaveEqn}
\end{equation}
In order to determine the propagation of the waves we insert the
plane wave ansatz $\mbox{\boldmath$E$} = {\mbox{\boldmath$E$}}^0
e^{i k_\mu x^\mu} = {\mbox{\boldmath$E$}}^0 e^{- i
(\mbox{\boldmath$\scriptstyle k$} \cdot
\mbox{\boldmath$\scriptstyle x$} - \omega t)}$ into wave equation
and take the derivatives of the amplitude into account
\begin{eqnarray}
0 & = & \ddot E^0_i - 2 i \omega \dot E^0_i - \omega^2 E^0_i
- \left(\Delta E^0_i + 2 i k \cdot \nabla E^0_i - k^2 E^0_i\right)
+ \partial_i \partial_j E^0_j + i k_j \partial_i E^0_j + i k_i \partial_j E^0_j
- k_j E^0_j k_i \nonumber\\
& & + 2 \chi^{i\mu\nu j} \left(\partial_\mu \partial_\nu E^0_j
+ i k_\mu \partial_\nu E^0_j + i k_\nu \partial_\mu E^0_j - k_\mu k_\nu\right)
+ 2 i \omega \chi^{i0j} E^0_j - i k_k \chi^{ikj} E^0_j
\end{eqnarray}
To first order the amplitude is constant, $\partial E^0_i = 0$.
The corresponding real part gives the relation
\begin{equation}
0 = \left((\omega^2 - {\mbox{\boldmath$k$}}^2)\delta_{ij} + k_i k_j
+ 2 \chi^{i \mu j \nu} k_\mu k_\nu\right) E_j^0 \, . \label{0thorder}
\end{equation}
The next order is related to the derivatives of the amplitude and
gives, due to the real and imaginary part, two equations
\begin{eqnarray}
0 & = & - 2 \omega \dot E^0_i - 2 k \cdot \nabla E^0_i
+ k_j \partial_i E^0_j + k_i \partial_j E^0_j \nonumber\\
& & + 2 \chi^{i\mu\nu j} \left(k_\mu \partial_\nu E^0_j
+ k_\nu \partial_\mu E^0_j\right) + 2 \omega \chi^{i0j} E^0_j
- k_k \chi^{ikj} E^0_j \label{approx1} \\
0 & = & \ddot E^0_i - \Delta E^0_i + \partial_i \partial_j E^0_j
+ 2 \chi^{i\mu\nu j} \partial_\mu \partial_\nu E^0_j \, .
\end{eqnarray}

\subsection{The dispersion relation}

The existence of a solution $E^0_j$ for Eq.(\ref{0thorder})
requires that the determinant of the coefficient matrix vanishes
\begin{equation}
0 = \det\left((\omega^2 - {\mbox{\boldmath$k$}}^2)\delta_{ij} + k_i k_j
+ 2 \chi^{i \mu j \nu} k_\mu k_\nu\right) \, ,
\end{equation}
which establishes a relation between the frequency $\omega$ and
the wave vector $\mbox{\boldmath$k$}$. To first order in the
modifications, the frequency is given by
\begin{equation}
\omega = \left(1 + \rho(k) \pm \sqrt{\sigma^2(k) - \rho^2(k)}\right) |\mbox{\boldmath$k$}|
\end{equation}
with
\begin{eqnarray}
\rho & = & \frac{1}{2} \left(\eta_{\rho\sigma} \chi^{\rho \mu \sigma \nu}
- \chi^{\mu \nu 0 i} n_i\right) n_\mu n_\nu \\
\sigma^2 & = & \frac{1}{2} \left(\chi^{i\mu 0 \nu} \chi^{j \rho 0 \sigma} n_i n_j
+ 2 \chi^{i \mu 0 \nu} \chi^{j \rho i \sigma} n_j
+ \chi^{i\mu j\nu} \chi^{j\rho i\sigma}\right) n_\mu n_\nu n_\rho n_\sigma \, ,
\end{eqnarray}
with $n_\mu = k_\mu/\omega = (1,
\mbox{\boldmath$k$}/|\mbox{\boldmath$k$}|)$. This generalizes the
results in \cite{Ni74,KosteleckyMewes02}. As mentioned in
\cite{KosteleckyMewes02}, the velocity of light is, to leading
order in the anomalous terms, given by $v = 1 + \rho(k) \pm
\sqrt{\sigma^2 - \rho^2}$.

It is possible to simplify further the above quantities $\rho$ and
$\sigma$. Inserting the irreducible decomposition, we get for
$\rho$
\begin{equation}
\rho = - \frac{1}{2} \Phi^{\mu\nu} n_\mu n_\nu \, .
\end{equation}
The $\sigma$--term can be considerably simplified by using the
condition $\chi^{(\alpha\beta\mu)\nu} = 0$ from the uniqueness of
the Cauchy problem. For doing so, we complete the summation over
$i$, $j$, etc. to a summation over all indices, e.g.,
$\chi^{i\mu\nu 0} n_i n_\mu n_\nu = \chi^{\alpha\mu\nu 0} n_\alpha
n_\mu n_\nu - \chi^{0\mu\nu 0} n_\mu n_\nu =  - \chi^{0\mu\nu 0}
n_\mu n_\nu$. Finally we get
\begin{equation}
\sigma^2 = \frac{1}{2} \eta_{\alpha\gamma} \eta_{\beta\delta}
\chi^{\alpha\mu\nu\beta} \chi^{\delta\rho\sigma\gamma} n_\mu n_\nu n_\rho n_\sigma \, ,
\end{equation}
where we set $\eta^{\mu\nu} n_\mu n_\nu = 0$ since this produces
higher order corrections only.

\subsection{No birefringence}

Kosteleck\'{y} and Mewes
\cite{KosteleckyMewes01,KosteleckyMewes02} analyzed very carefully
the light from distant galaxies and inferred that to very high
precision there is no birefringence. That means $0 = \sigma^2 -
\rho^2$, that is
\begin{equation}
0 = \frac{1}{2} \left(\chi^{\alpha\mu\nu}{}_\beta \chi^{\beta\rho\sigma}{}_\alpha
- \frac{1}{2} \Phi^{\mu\nu} \Phi^{\rho\sigma} \right) n_\mu n_\nu n_\rho n_\sigma \, .
\end{equation}
This means that the totally symmetric part of
$\chi^{\alpha\mu\nu}{}_\beta \chi^{\beta\rho\sigma}{}_\alpha -
\frac{1}{2} \Phi^{\mu\nu} \Phi^{\rho\sigma}$ has to be
proportional to the unperturbed metric $\eta^{\mu\nu}$:
\begin{equation}
\chi^{\alpha(\mu\nu}{}_\beta \chi^{|\beta|\rho\sigma)}{}_\alpha
- \frac{1}{2} \Phi^{(\mu\nu} \Phi^{\rho\sigma)}
= \eta^{(\mu\nu} \mu^{\rho\sigma)} \label{nobirefringence}
\end{equation}
where $\mu$ is some symmetric tensor. Using again the
decomposition (\ref{ctcphm}) we get after some lengthy
calculations
\begin{equation}
{}^{(1)}W^{\mu\nu\rho\sigma} = 0 \, , \qquad \Psi^{\mu\nu} = 0 \, .
\end{equation}
With this result, the constitutive tensor reduces to
\begin{equation}
\chi^{\alpha\beta\mu\nu} = - \eta^{\mu[\alpha} \Phi^{\beta]\nu} +
\eta^{\nu[\alpha} \Phi^{\beta]\mu} - \frac{1}{2} \eta^{\alpha[\mu}
Z^{\nu]\beta} + \frac{3}{2} \eta^{\beta[\mu} Z^{\nu]\alpha} +
\frac{1}{2} \eta^{\alpha\beta} Z^{\mu\nu} - \frac{1}{6} W
\eta^{\alpha [\mu} \eta^{|\beta|\nu]} \label{ctcphmnb1}
\end{equation}
and the Maxwell equations are
\begin{eqnarray}
4 \pi j^\alpha & = & \partial_\nu F^{\mu\nu}
- 2 \eta^{\mu[\alpha} \Phi^{\beta]\nu} \partial_\beta F_{\mu\nu} \nonumber\\
& & - \frac{1}{2} \eta^{\alpha\mu} Z^{\nu\beta} \partial_\beta F_{\mu\nu}
+ \frac{3}{2} \eta^{\beta\mu} Z^{\nu\alpha} \partial_\beta F_{\mu\nu}
+ \frac{1}{2} \eta^{\alpha\beta} Z^{\mu\nu} \partial_\beta F_{\mu\nu}
- \frac{1}{6} W \eta^{\alpha \mu} \eta^{\beta\nu} \partial_\beta F_{\mu\nu} \, . \label{GME3}
\end{eqnarray}
These are the most general Maxwell equations which do not lead to birefringence.

\subsection{Isotropy of speed of light}

If we add experiments on the isotropy of light propagation, that
is, Michelson--Morley type experiments either based on
interferometers or cavities (see, e.g. \cite{Muelleretal03c}),
then we have access to the additional tensor $\Phi^{\mu\nu}$ only:
If the speed of light is required to be isotropic, then we get the
condition
\begin{equation}
\Phi^{\mu\nu} = 0
\end{equation}
leading to the constitutive tensor
\begin{equation}
\chi^{\alpha\beta\mu\nu} = - \frac{1}{2} \eta^{\alpha[\mu} Z^{\nu]\beta}
+ \frac{3}{2} \eta^{\beta[\mu} Z^{\nu]\alpha}
+ \frac{1}{2} \eta^{\alpha\beta} Z^{\mu\nu}
- \frac{1}{6} W \eta^{\alpha [\mu} \eta^{|\beta|\nu]} \label{ctcphmnb}
\end{equation}
and the Maxwell equations
\begin{eqnarray}
4 \pi j^\alpha & = & \left(1 - {\textstyle \frac{1}{6}} W\right)
\partial_\nu F^{\mu\nu} - \frac{1}{2} \eta^{\alpha\mu}
Z^{\nu\beta} \partial_\beta F_{\mu\nu} + \frac{3}{2}
\eta^{\beta\mu} Z^{\nu\alpha} \partial_\beta F_{\mu\nu} +
\frac{1}{2} \eta^{\alpha\beta} Z^{\mu\nu} \partial_\beta
F_{\mu\nu} \, . \label{GME4}
\end{eqnarray}
That means, vanishing birefringence and isotropy of light
propagation is not enough to establish the Lorentz invariance of
the theory -- in contrast to the Lagrangian based SME where the
constitutive tensor is of ordinary form only. Our frame is much
more general than a Lagrangian based theory. Since the
$Z^{\mu\nu}$ cannot be probed by radiation phenomena we later
analyze the field created by point charges and magnetic moments.
But first we consider the propagation of the polarization of the
radiation field.

\subsection{Propagation of polarization states}

In order to determine the propagation of the amplitude from the
wave equation we reformulate (\ref{approx1}) and get
\begin{eqnarray}
0 & = & \left(\delta_i^j - 2 \chi^{i00j} - 2 \chi^{i(0k)j} n_k\right) \dot E^0_j
+ \left(n_l \delta_i^j - \frac{1}{2} n_j \delta_i^l - \frac{1}{2} n_i \delta^{lj}
- 2 \chi^{i(0l)j} - 2 \chi^{i(kl)j} n_k\right) \partial_l E^0_j \nonumber\\
& & + \left(2 \omega \chi^{i0j} - k_k \chi^{ikj}\right) E^0_j
\label{PropE1}
\end{eqnarray}

We now apply the theorem from matrix theory \cite{Gantmakher59}
that for any matrix $A$ there is a minor $M_A$ so that $M_A A =
\det A$. (In the case that $\det A$ possesses zeros of higher
order, then $M_A$ can be chosen so that on the right hand side
there appears a first order zero only: $M^\prime_A A = \det^\prime
A$. In the unperturbed case $\det A = \omega^2 (\omega^2 -
k^2)^2$, so that $\det^\prime A = \omega^2 (\omega^2 - k^2)$). In
our case we have $\det A = \omega^2 (\omega_+^2 - k^2) (\omega_-^2
+ k^2)$ with first order zeros only. Differentiation of $M_A A =
\det A$ with respect to the wave vector gives
\begin{equation}
\frac{\partial (M_A)_i{}^l}{\partial k_\mu} A_l{}^j
+ (M_A)_i{}^l \frac{\partial A_l{}^j}{\partial k_\mu}
= \frac{\partial \det A}{\partial k_\mu} \delta^j_i = v^\mu \delta^j_i
\end{equation}
where $v^\mu$ is the group velocity.

We apply this result to the amplitude. On--shell, the first term
will vanish due to $A_i{}^j E^0_j = 0$.  Therefore, we have
on--shell
\begin{equation}
(M_A)_i{}^l \frac{\partial A_l{}^j}{\partial k_\mu} \partial_\mu E^0_j
= v^\mu \delta^j_i  \partial_m E^0_j \, .
\end{equation}
In our case
\begin{eqnarray}
(M_A)_i{}^l \frac{\partial A_l{}^j}{\partial k_\mu} \partial_\mu E^0_j
& = & (M^\prime_A)_i{}^l \frac{\partial}{\partial k_\mu} \left((\omega^2 - k^2)
\delta_l^j + k^l k_j - 2 \chi^{l\rho\sigma j} k_\rho k_\sigma\right)
\partial_\mu E^0_j \nonumber\\
& = & (M^\prime_A)_i{}^l \left(2 k^\mu \delta_l^j + \delta^{l\mu} k_j
+ k^l \delta^\mu_j - 4 \chi^{l(\mu\sigma) j} k_\sigma\right) \partial_\mu E^0_j \, ,
\end{eqnarray}
where on the right hand side the matrix in (\ref{PropE1}) shows
up. Therefore, the amplitude precesses during the transport along
the light rays
\begin{equation}
v^\mu \partial_\mu E_i^0 = - \left(2 \omega \chi^{i0j} - k_k \chi^{ikj}\right) E^0_j \, .
\end{equation}
The polarization is sensitive only to the mass tensor
$\chi^{\mu\rho\sigma}$, no additional information for the
constitutive tensor can be deduced.
%Indeed, a precession of the polarization is always
%encoded in connection--like terms.
%These are included in the $\chi^{\alpha\mu\nu}$.
The axion part of $\chi^{\alpha\mu\nu}$ is an example leading to a
non--vanishing evolution of $E^0_i$ which has been analyzed in
\cite{CarrollFieldJackiw90,CarrollField91}.

In terms of the irreducible decomposition (\ref{irredmass}) we get
\begin{equation}
v^\mu \partial_\mu E_i^0 = - \omega \left(2 ({}^{(1)}\chi^{i0j}
+ \epsilon^{i0jk} a_k + \eta^{i[0} t^{j]}) - n_k ({}^{(1)}\chi^{ikj}
+ \epsilon^{ikj0} a_0 + \eta^{i[k} t^{j]})\right) E^0_j \, .
\end{equation}
To first order in the parameters encoded in
$\chi^{\mu\rho\sigma}$, we get for the two characteristic
projections
\begin{eqnarray}
n_i v^\mu \partial_\mu E_i^0 & = & - \omega \left(2 \left({}^{(1)}\chi^{i0j} n_i E^0_j
+ \epsilon^{i0jk} n_i a_k E^0_j\right) - \left({}^{(1)}\chi^{ikj} n_i n_k E^0_j
- \frac{1}{2} t^{j} E^0_j\right)\right) \\
E^0_i v^\mu \partial_\mu E_i^0 & = & - \omega \left(2 \left({}^{(1)}\chi^{i0j}
E^0_i E^0_j + \frac{1}{2} t^{0} (E^0)^2\right) - \left({}^{(1)}\chi^{ikj} n_k E^0_i E^0_j
+ \frac{1}{2} n_k t^{k} (E^0)^2\right)\right)
\end{eqnarray}
where we used $k_i E_i = {\cal O}(\chi)$. Each part can be
isolated by varying independently the polarization $E^0_i$ and the
direction of propagation. If we assume that no precession of the
polarization will be observed (what is well confirmed by
observations, see below), then $\chi^{\mu\rho\sigma}$ has to
vanish. Since $\chi^{\mu\rho\sigma}$ is related to CNC,
observations of the precession of the polarization of
electromagnetic radiation can be used for testing the validity of
charge conservation.

\section{3+1 decomposition} \label{PointCharges}

For a further analysis of our generalized Maxwell equation we
perform a 3+1 decomposition. With the two 3--vecors  $\zeta^i :=
\frac{3}{2} Z^{0i}$ and $\hat\zeta_i := \frac{3}{4} \epsilon_{ijk}
Z^{jk}$ we get from (\ref{GME4}) in SI units
\begin{align}
\frac{\rho}{\epsilon_0} & = - \left(1 - {\textstyle \frac{1}{6}} W\right)
\mbox{\boldmath$\nabla$} \cdot \mbox{\boldmath$E$}
- \hat{\mbox{\boldmath$\zeta$}} \cdot (\mbox{\boldmath$\nabla$} \times \mbox{\boldmath$E$})
- c \mbox{\boldmath$\zeta$} \cdot (\mbox{\boldmath$\nabla$} \times \mbox{\boldmath$B$})
\label{GenMax3p11} \\
\mu_0 \mbox{\boldmath$j$} & = \frac{1}{c^2} \left(1 - {\textstyle \frac{1}{6}} W\right)
\dot{\mbox{\boldmath$E$}} - \frac{1}{c^2} \hat{\mbox{\boldmath$\zeta$}} \times
\dot{\mbox{\boldmath$E$}} - (\hat{\mbox{\boldmath$\zeta$}} \cdot \mbox{\boldmath$\nabla$})
\mbox{\boldmath$B$} + \left(1 - {\textstyle \frac{1}{6}} W\right) \mbox{\boldmath$\nabla$}
\times \mbox{\boldmath$B$}
- \frac{1}{c} \mbox{\boldmath$\nabla$} (\mbox{\boldmath$\zeta$} \cdot \mbox{\boldmath$E$})
+ \frac{1}{c} \mbox{\boldmath$\zeta$} (\mbox{\boldmath$\nabla$} \cdot \mbox{\boldmath$E$})
\, , \label{GenMax3p12}
\end{align}
where $c^2 = 1/(\epsilon_0 \mu_0)$. This kind of violation of
Lorentz invariance in the Maxwell theory encoded in the two
parameters $\mbox{\boldmath$\zeta$}$ and
$\hat{\mbox{\boldmath$\zeta$}}$ has not been treated hitherto. It
can be shown explicitly that this is compatible with all the
requirements stated until now. The remaining Lorentz--invariance
violating terms $\mbox{\boldmath$\zeta$}$ and
$\hat{\mbox{\boldmath$\zeta$}}$ can be probed by studying the
fields of point charges and magnetic moments only. These
coefficients cannot be probed by radiation phenomena. Here we
restrict ourselves to the principal part of the Maxwell equations
only. Effects due to a particular choice for the mass term have
been discussed in \cite{LaemmerzahlHaugan01} where an anisotropic
speed of light and CNC has been derived; similar effects will
occur for the general case.

\subsection{Solution for a point charge}

The generalized Maxwell equations for a point charge at the origin
are given by (\ref{GenMax3p11},\ref{GenMax3p12}) with $\rho = q
\delta(r)$ and $\mbox{\boldmath$j$} = 0$. Since we have a static
problem, we neglect the time derivatives. We furthermore chose
$\mbox{\boldmath$E$} = \mbox{\boldmath$\nabla$} \phi$ and
$\mbox{\boldmath$B$} = \mbox{\boldmath$\nabla$} \times
\mbox{\boldmath$A$}$ and the gauge $\mbox{\boldmath$\nabla$} \cdot
\mbox{\boldmath$A$} = 0$. Then the generalized Maxwell equations
are
\begin{align}
\frac{q}{\epsilon_0} \delta(r) & = - \left(1 - {\textstyle \frac{1}{6}}
W\right) \Delta\phi + c \mbox{\boldmath$\zeta$} \cdot \Delta \mbox{\boldmath$A$} \\
0 & = - (\hat{\mbox{\boldmath$\zeta$}} \cdot \mbox{\boldmath$\nabla$})
\mbox{\boldmath$\nabla$} \times \mbox{\boldmath$A$}
- \left(1 - {\textstyle \frac{1}{6}} W\right) \Delta \mbox{\boldmath$A$}
- \frac{1}{c} \mbox{\boldmath$\nabla$} (\mbox{\boldmath$\zeta$}
\cdot \mbox{\boldmath$\nabla$} \phi) + \frac{1}{c} \mbox{\boldmath$\zeta$}
\Delta\phi \, . \label{pointcharge2}
\end{align}

\begin{figure}[t]
\begin{center}
\includegraphics*[scale=0.5]{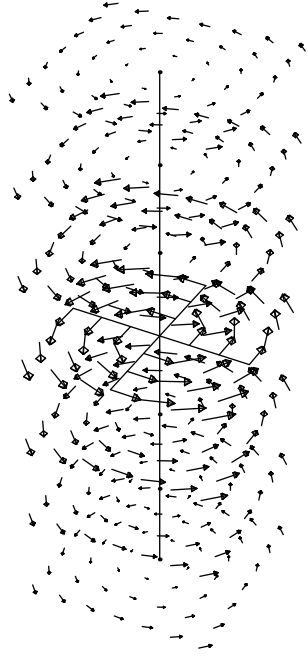}
\end{center}
\caption{The magnetic field of a point charge located at the
origin. The vector field $\mbox{\boldmath$\zeta$}$ is assumed to
point in the $z$--direction. \label{MagneticField}}
\end{figure}

The solution should be of the form
\begin{equation}
\phi = \frac{1}{4 \pi \epsilon_0} \frac{q}{r} + \delta\phi \, ,
\qquad \mbox{\boldmath$A$} = \delta\mbox{\boldmath$A$} \, ,  \label{PointChargeAnsatz}
\end{equation}
where $\delta\phi$ and $\delta\mbox{\boldmath$A$}$ are small
quantities, at most of the order of $\mbox{\boldmath$\zeta$}$.
Therefore,  modifications of the static electrical potential will
be of second order only, $\delta\phi = {\cal
O}({\mbox{\boldmath$\zeta$}}^2)$. However, inserting the
unperturbed solution $\phi = q/(4\pi \epsilon_0 r)$ into the
second equation (\ref{pointcharge2}), we get to first order in the
perturbations
\begin{equation}
\Delta \mbox{\boldmath$A$} = \frac{1}{c} \mbox{\boldmath$\zeta$} \Delta\phi
= - \frac{1}{c} \mbox{\boldmath$\zeta$} \frac{q}{\epsilon_0} \delta(r)
\end{equation}
with the solution
\begin{equation}
\mbox{\boldmath$A$} = \frac{q \mbox{\boldmath$\zeta$}}{4 \pi \epsilon_0 c r} \, .
\end{equation}
This gives a magnetic field
\begin{equation}
\mbox{\boldmath$B$} = \frac{q}{4 \pi \epsilon_0 c}
\frac{\mbox{\boldmath$\zeta$} \times \mbox{\boldmath$r$}}{r^3} \, . \label{zetamagnfield}
\end{equation}
Therefore, our model includes the feature that a point charge also
creates a magnetic field. This field is different from a field of
a magnetic moment. If the point charge is at the origin, and if we
take the coordinate system such that $\mbox{\boldmath$\zeta$}$
points in ${\mbox{\boldmath$e$}}_z$ direction, then the magnetic
field lines are circles in the $x$--$y$--plane, similar to the
magnetic field lines around a wire, see Fig.~\ref{MagneticField}.
The strength, however, varies with $1/r^2$ where $r$ is the
distance from the origin.

If we take a charged line with line--density $\lambda$ in
direction $\mbox{\boldmath$n$}$, then the magnetic field is
\begin{equation}
\mbox{\boldmath$B$} = - \frac{\lambda}{2\pi \epsilon_0 c} \frac{\mbox{\boldmath$\zeta$}
\cdot \mbox{\boldmath$n$}}{\rho} {\mbox{\boldmath$e$}}_\varphi
- \frac{\lambda}{2\pi \epsilon_0 c} \frac{\left(\mbox{\boldmath$\zeta$}
\times {\mbox{\boldmath$e$}}_\rho\right) \cdot \mbox{\boldmath$n$}}{\rho}
\mbox{\boldmath$n$} \, ,
\end{equation}
where $\rho$ is the distance from the charged line,
${\mbox{\boldmath$e$}}_\rho$ the radial unit vector orthogonal to
$\mbox{\boldmath$n$}$, and ${\mbox{\boldmath$e$}}_\varphi$ the
unit tangent vector of a circle around that line. For
$\mbox{\boldmath$n$} \sim \mbox{\boldmath$\zeta$}$ the magnetic
field is
\begin{equation}
\mbox{\boldmath$B$} = - \frac{\lambda}{2\pi \epsilon_0 c} \zeta
\frac{{\mbox{\boldmath$e$}}_\varphi}{\rho} \, .
\end{equation}
This has the form of the magnetic field of a wire carrying the
current $-\lambda\zeta/(2 \pi \epsilon_0 c)$. For
$\mbox{\boldmath$n$} \bot \mbox{\boldmath$\zeta$}$
\begin{equation}
\mbox{\boldmath$B$} = - \frac{\lambda}{2\pi \epsilon_0 c}
\frac{\mbox{\boldmath$n$} \cdot \left(\mbox{\boldmath$\zeta$}
\times {\mbox{\boldmath$e$}}_\rho\right)}{\rho} \mbox{\boldmath$n$} \, .
\end{equation}

\subsection{Solution for a magnetic moment}

If the source of the Maxwell equations is a magnetic moment
$\mbox{\boldmath$m$}$ localized at the origin, then the Maxwell
equations are  (\ref{GenMax3p11},\ref{GenMax3p12}) with $\rho = 0$
and $\mbox{\boldmath$j$} = \mbox{\boldmath$m$} \times
\mbox{\boldmath$\nabla$} \delta(r)$. We assume again a static
situation
\begin{align}
0 & = - \left(1 - {\textstyle \frac{1}{6}} W\right) \Delta\phi
+ c \mbox{\boldmath$\zeta$} \cdot \Delta \mbox{\boldmath$A$}  \\
\mu_0 \mbox{\boldmath$m$} \times \mbox{\boldmath$\nabla$}
\delta(r) & = - (\hat{\mbox{\boldmath$\zeta$}} \cdot \mbox{\boldmath$\nabla$})
\mbox{\boldmath$\nabla$} \times \mbox{\boldmath$A$} - \left(1
- {\textstyle \frac{1}{6}} W\right) \Delta \mbox{\boldmath$A$}
- \frac{1}{c} \mbox{\boldmath$\nabla$} (\mbox{\boldmath$\zeta$} \cdot
\mbox{\boldmath$\nabla$} \phi) + \frac{1}{c} \mbox{\boldmath$\zeta$} \Delta\phi
\end{align}
and make the ansatz
\begin{equation}
{\mbox{\boldmath$A$}} = \frac{\mu_0}{4 \pi}
\frac{\mbox{\boldmath$m$}\times \mbox{\boldmath$r$}}{r^3} + \delta
\mbox{\boldmath$A$} \, , \qquad \phi = \delta\phi \, .
\end{equation}
Similar to the previous case, insertion of the unperturbed
solution for $\mbox{\boldmath$A$}$ into the first equation leads,
in first order of the perturbations, to
\begin{equation}
\Delta\phi = c \mbox{\boldmath$\zeta$} \cdot \left(\mu_0 \mbox{\boldmath$m$}
\times \mbox{\boldmath$\nabla$} \delta(r)\right) = \mu_0 c \left(\mbox{\boldmath$\zeta$}
\times \mbox{\boldmath$m$} \right) \mbox{\boldmath$\nabla$} \delta(r) \, .
\end{equation}
This is the equation for an electrical dipole with dipole moment
$\mbox{\boldmath$d$} = \mu_0 c \epsilon_0 \mbox{\boldmath$\zeta$}
\times \mbox{\boldmath$m$}$. The solution is
\begin{equation}
\phi = \frac{\mu_0 c}{4 \pi} \frac{\left(\mbox{\boldmath$\zeta$}
\times \mbox{\boldmath$m$}\right) \cdot \mbox{\boldmath$r$}}{r^3} \, .
\end{equation}
Therefore, a magnetic moment also creates an electric field.
This feature is ''dual'' to the previous case.

The parameter $\hat{\mbox{\boldmath$\zeta$}}$ gives rise to small
deviations from the unperturbed quantities only, that is, it
induces a small additional term $\delta \phi$ for a given $\phi =
q/(4\pi\epsilon_0 r)$ and a small additional
$\delta\mbox{\boldmath$A$}$ for a large $\mbox{\boldmath$A$} =
\mu_0 \mbox{\boldmath$m$} \times \mbox{\boldmath$r$}/(4 \pi r^3)$.
This cannot be measured such precisely. On the contrary,
measurements of the parameter $\mbox{\boldmath$\zeta$}$ always
amount to a null--test and are, thus, much more precise.

\section{Confrontation with experiment}

Using astrophysical observations and laboratory experiments, the
possibility of birefringence and an anisotropy of the velocity of
light has been estimated to very hight precision. Kosteleck\'{y}
and Mewes gave an upper limit of birefringence by
$|{}^{(1)}W^{\mu\nu\rho\sigma}|, |\Psi^{\mu\nu}| \leq 10^{-32}$
\cite{KosteleckyMewes02}. The today's most precise
Michelson--Morely experiment by M\"uller and coworkers
\cite{Muelleretal03c} using optical resonators restricted a
possible anisotropy to $|\Phi^{\mu\nu}| \leq 10^{-15}$. Here, all
estimates are valid for each component in the actual laboratory
frame. See \cite{Muelleretal03,Muelleretal03d} for a refined
description of that experiments.

A possible precession of the polarization has also been estimated
from astrophysical observations to very high precision. In a model
with a totally antisymmetric $\chi^{\mu\rho \sigma}$ Carroll,
Field and Jackiw analyzed the polarization from distant galaxies
\cite{CarrollFieldJackiw90} and obtained, since no precession of
the polarization has been found, the estimate
$\chi^{[\mu\rho\sigma]} \leq 10^{-42}\;\hbox{GeV}$ which is
equivalent to $|\chi^{[\mu\rho\sigma]}| \leq 3 \cdot
10^{-17}\;{\hbox{s}}^{-1}$. Since the other irreducible parts of
$\chi^{\mu\rho\sigma}$ lead to a precession of the polarization,
too, we extend this result to the other parts:
$|\chi^{\mu\rho\sigma}| \leq 3 \cdot 10^{-17}\;{\hbox{s}}^{-1}$.
This in particular also means that the charge is conserved to that
order: $|\dot Q/Q| \leq 3 \cdot 10^{-17}\;{\hbox{s}}^{-1}$. Since
this result is not connected with the choice of a time dependence
of any other ''constant'', it represents a clear and dedicated
statement about charge conservation. This result on the
conservation of the electric charge is one order better than what
one gets from tests of the time--dependence of the fine structure
constant \cite{Marionetal03} by assuming a constant $c$ and
$\hbar$.

As far as the new Lorentz invariance violating parameters
$\mbox{\boldmath$\zeta$}$ and $\hat{\mbox{\boldmath$\zeta$}}$ are
concerned, we are, unfortunately, not aware of any dedicated
experiment searching for, e.g., a magnetic field which is created
by a point charge. In order to get some feeling for the accuracy
of the validity of the ordinary Maxwell equations, that is for
$\mbox{\boldmath$\zeta$} = 0$ and $\hat{\mbox{\boldmath$\zeta$}} =
0$, we discuss the accuracy of some possible high precision
measurements of magnetic fields. Magnetic field can be measured
with the help of SQUIDs (measurements based on the Hall effect are
not such precise).

With SQUIDs weak magnetic fields of down to $10^{-14}\;\hbox{T}$
can be measured. We assume that even a dedicated search for a
magnetic field from a point charge does not lead to any magnetic
field larger than the SQUID sensitivity. Then, from $|\lambda
\zeta / (2 \pi \epsilon_0 c \rho)| \leq 10^{-14}\;\hbox{T}$ for a
line charge density\footnote{For negative charges, this line
charge density is the principal limit for a wire of 1 mm diameter
when taking into account that field emission starts at approx.
$10^{11}\;\hbox{V/m}$ at the surface of the wire. Also the
possibilities to create a sufficient high voltages limits the
charge line density. Therefore we extend this limit to positive
charges, too.} $\lambda = 0.01 \;\hbox{C/m}$ at a distance of 1
cm, we get the estimate $|\zeta| \leq 2.7 \cdot 10^{-17}$.
However, this is just the estimate which would result if such a
kind of experiments yields a null--result; a dedicated experiment
of this kind has not yet been carried through.

Another method to search for this kind of effects is to use atomic
spectroscopy. Since the charge of the proton leads to a magnetic
field, a hyperfine splitting, additional to the usual one, should
occur. Due to the different radial structure of the magnetic
field, the result also should be different from the ordinary
hyperfine splitting. With obvious notations, we get for the
interaction Hamiltonian of an electron in the magnetic field
(\ref{zetamagnfield}) of the nucleus
\begin{eqnarray}
H_{\zeta} & = & {\mbox{\boldmath$\mu$}}_{\rm el} \cdot
\frac{q \mbox{\boldmath$\zeta$} \times \mbox{\boldmath$r$}}{4 \pi \epsilon_0 c r^3} \, .
\end{eqnarray}
If we choose the $z$--axis in direction of
$\mbox{\boldmath$\zeta$}$, then the corresponding energy shift
$\Delta E_{nlm} = \langle \psi_{nlm} \mid H_\zeta \mid
\psi_{nlm}\rangle$ is
\begin{equation}
\Delta E_{nlm} %= \frac{q \zeta_z}{4 \pi \epsilon_0 c}
%\int \psi^*_{nlm} \left(\mu_y \sin\vartheta \cos\varphi
%- \mu_x \cos\vartheta\right) \psi_{nlm} dr
%\sin\vartheta d\vartheta d\varphi
= - \frac{q \zeta_z \mu_x}{4 \pi \epsilon_0 c}
\int \psi^*_{nlm} \cos\vartheta \psi_{nlm} dr \sin\vartheta d\vartheta d\varphi \, .
\end{equation}
This does not vanish for, e.g., $\psi_{210} = R_{21} Y_{10}$ where
$R_{21} = \frac{1}{\sqrt{3}} \frac{1}{(2 a)^{3/2}} \frac{r}{a}
e^{-r/(2a)}$, $Y_{10} = \sqrt{\frac{3}{4\pi}} \cos\vartheta$ where
$a$ in the Bohr radius (contrary to the ordinary hyperfine
splitting, there is no shift for the $s$ states). In this case we
get
\begin{equation}
\Delta E_{210} %= - \frac{q \zeta_z \mu_x}{4 \pi \epsilon_0 c}
%\int R_{21}^2 dr \int (Y_{10})^2
%\cos\vartheta \sin\vartheta d\vartheta \int d\varphi
%= - \frac{q \zeta_z \mu_x}{4 \pi \epsilon_0 c}
%\frac{1}{12 a^2} \frac{1}{2\pi} 2 \pi
= - \frac{q \zeta_z \mu_x}{48 \pi \epsilon_0 c a^2} \, .
\end{equation}
With $\mu_z = e \hbar/m_e$ this yields $\Delta E_{210} = \zeta_z
1.8 \cdot 10^{-2}\;\hbox{eV}$. The state of the art of high
precision measurements of energy levels is of the order $\Delta
E/E \approx 10^{-15}$. Since the measured energy levels are still
well described within the standard theory one gets for energies of
about 10 eV at best an estimate $|\zeta_z| \leq 10^{-14}$ which,
however, is not as good as a direct measurement discussed above
might yield.

\section{Conclusion}

We discussed the most general model for the dynamics of the
electromagnetic field which is linear and of first order in the
derivative of the field strength. This is tantamount to the most
general ansatz linear in the field strength leading to violation
of Lorentz invariance in the Maxwell theory. It is shown that the
condition of dynamical consistency is responsible for the charge
conservation of the principal part of the differential equations:
CNC can be induced only by an additional term without derivative
in generalized Maxwell equations. Our model cannot be derived from
a Lagrangian. Due to this very general and systematic approach we
were able to identify CNC terms and, furthermore, Lorentz
violating terms encoded in the constitutive tensor which are
beyond the SME. Beside the usual tests of the time--dependence of
the fine structure constant (which interpretation depends on
assumptions on the time--dependence of the velocity of light and
the Planck's constant), the CNC parameters can be tested through
its effects on the polarization of electromagnetic radiation.
Using a previous analysis of astrophysical observations on the
polarization of the light of distant galaxies, charge conservation
can be confirmed at a slightly better level than from laboratory
$\dot\alpha$--experiments.

If we describe all experiments related to CNC in terms of the
parameters ${}^{(1)}\chi^{\mu\rho\sigma}$ and $t^\mu$, then the
results from electron disappearing experiments, from searches for
a time--dependence of the fine structure constant and from
astrophysical observations on a precession of the polarization of
light from distant galaxies can be compared, see Table
\ref{CNCcomparison}. Being a discrete process, the electron
disappearing in principle may have a different cause than a
continuous variation of the elementary charge. Therefore, the
first experiment in Table \ref{CNCcomparison} has a slightly
different status than the other two.

\begin{table}
\begin{tabular}{|l|c|}\hline
Experiment/observation & estimate on charge non--conservation parameter \\ \hline
electron disappearing \cite{Steinbergetal75} & $6.3 \cdot 10^{-30} \;{\hbox{s}}^{-1}$ \\
time--dependence of fine structure constant \cite{Marionetal03}
& $1.2 \cdot 10^{-23}\;{\hbox{s}}^{-1}$ \\
precession of polarization \cite{CarrollFieldJackiw90}
& $3 \cdot 10^{-17}\;{\hbox{s}}^{-1}$\\ \hline
\end{tabular}
\caption{Experimental constraints on the charge non--conservation
 parameters ${}^{(1)}\chi^{\mu\rho\sigma}$ and $t^\mu$ of
Eq.(\ref{cnc1}) from various experiments and observations.
\label{CNCcomparison}}
\end{table}

The Lorentz invariance violating terms in the constitutive tensor
which are beyond the SME cannot be probed by radiation phenomena.
This means, optical experiments (in particular Michelson--Morley
experiments together with observations on birefringence) are not
sufficient to prove or to establish uniquely the Lorentz
invariance of the theory. These additional Lorentz violating terms
lead to the effects that an electrical point charge, beside the
ordinary Coulomb potential, also creates a magnetic field and a
magnetic moment also an electric field. A rough discussion of what
might be expected by carrying through dedicated experiments using
current technology leads to estimates of
$|\mbox{\boldmath$\zeta$}| \leq 2 \cdot 10^{-17}$. Therefore
experiments which explore the fields of point charges and
point--like magnetic moments are needed and we strongly suggest
experimentalists to carry out such experiments.

\begin{acknowledgments}
We like to thank H. Dittus, F.W. Hehl, and A. Kosteleck\'{y} for
fruitful discussions. Financial support from the German Academic
Exchange Service DAAD, the German Space Agency DLR, the Alexander
von Humboldt--Stiftung, and CONACYT Grant 42191-F is acknowledged.
\end{acknowledgments}

\appendix

\section{Appendix: Decomposition of generalized constitutive tensor}

We shortly describe the irreducible decomposition of generalized
constitutive tensor $\chi^{\alpha\beta\mu\nu} =
\chi^{\alpha\beta[\mu\nu]}$ after Hehl et al. \cite{Hehletal95}.
First we split the tensor into its symmetric and antisymmetric
pieces
\begin{equation}
\chi^{\alpha\beta\mu\nu} = W^{\alpha\beta\mu\nu} + Z^{\alpha\beta\mu\nu}
\end{equation}
with
\begin{equation}
W^{\alpha\beta\mu\nu} := \chi^{[\alpha\beta]\mu\nu}\, ,
\qquad Z^{\alpha\beta\mu\nu} := \chi^{(\alpha\beta)\mu\nu} \, .
\end{equation}
Here, $R$ possesses 36 independent components and $Z$ 60
components. In the following, indices are raised and lowered with
the metric $\eta_{\mu\nu}$ and its inverse $\eta^{\mu\nu}$. We
define
\begin{eqnarray}
W & := & \chi_{\mu\nu}{}^{\mu\nu}\\
W^\alpha{}_\mu & := & W^{\alpha\beta}{}_{\beta\mu} \\
X^\alpha{}_\mu & := & - \frac{1}{6} \epsilon_\mu{}^{\tau\rho\sigma} W^{\alpha}{}_{\tau\rho\sigma} \\
X & := & - \frac{1}{6} \epsilon^{\mu\nu\rho\sigma} W_{\mu\nu\rho\sigma} \\
\Psi_{\alpha\mu} & := & X_{\alpha\mu} - \frac{1}{4} \eta_{\alpha\mu} X - X_{[\alpha\mu]}\\
\Phi_{\alpha\mu} & := & W_{\alpha\mu} - \frac{1}{4} \eta_{\alpha\mu} W - W_{[\alpha\mu]} \\
Z^{\mu\nu} & := & \eta_{\alpha\beta} Z^{\alpha\beta\mu\nu} \\
Z^{\alpha\beta\mu\nu}_{\rm tracefree} & := & Z^{\alpha\beta\mu\nu} - \frac{1}{4} \eta^{\alpha\beta} Z^{\mu\nu} \\
Z^{\alpha\mu}_{\rm tracefree} & := & \eta_{\beta\nu} Z^{\alpha\beta\nu\mu}_{\rm tracefree} \\
\Delta^{\mu\nu} & := & Z^{[\mu\nu]}_{\rm tracefree} \\
Y^{\alpha\mu} & := & \frac{1}{6} \epsilon^\mu{}_{\rho\sigma\tau} Z^{\alpha\tau\rho\sigma}_{\rm tracefree}  \\
\Xi^{\alpha\mu} & := & Z^{\alpha\mu}_{\rm tracefree} - Z^{[\alpha\mu]}_{\rm tracefree} \\
\Upsilon^{\alpha\mu} & := & Y^{\alpha\mu} - Y^{[\alpha\mu]}
\end{eqnarray}
These tensors have the following interpretation

\bigskip
\begin{tabular}{lll}
${}^{(1)}W^{\alpha\beta\mu\nu}$ & Weyl tensor & 10 components \\
$W$ & scalar of antisymmetric part  & 1 components \\
$W^\alpha{}_\mu$ & Ricci--tensor & 16 components \\
$X^\alpha_\mu$ & trace of the right--dual $\epsilon^{\mu\nu\rho\sigma} W_{\alpha\beta\rho\sigma} $ & 16 components \\
$X$ & pseudoscalar & 1 components \\
$\Psi^{\alpha\mu}$ & symmetric tracefree part of trace of right--dual & 9 components \\
$\Phi^{\alpha\mu}$ & symmetric tracefree part of Ricci--tensor & 9 components \\
${}^{(1)}Z^{\alpha\beta\mu\nu}$ & & 30 components \\
$Z_{\rm tracefree}^{\alpha\mu}$ & Ricci tensor of symmetric tracefree part, is trecefree & 9 components \\
$Z^{\mu\nu}$ & genuine trace of symmetric part (trace with respect to first two indices) & 6 components \\
$Y^{\alpha\mu}$ & trace of right--dual, is tracefree & 15 components \\
$\Upsilon^{\alpha\mu}$ & symmetric part of $Y^{\alpha\mu}$, tracefree & 9 components \\
$\Xi^{\alpha\mu}$ & symmetric part of $Z_{\rm tracefree}^{\alpha\mu}$, tracefree & 9 components \\
$\Delta^{\mu\nu}$ & antisymmetric part of Ricci tensor of symmetric tracefree part & 6 components \\
\end{tabular}

\bigskip
The irreducible decomposition of $\chi^{\alpha\beta\mu\nu}$ is then given by
\begin{equation}
\chi^{\alpha\beta\mu\nu} = \sum_{i = 1}^6 {}^{(i)}W^{\alpha\beta\mu\nu} + \sum_{i = 1}^5 {}^{(i)}Z^{\alpha\beta\mu\nu}
\end{equation}
with
\begin{eqnarray}
{}^{(1)}W^{\alpha\beta\mu\nu} & = & W^{\alpha\beta\mu\nu} - \sum_{i = 2}^6 {}^{(i)}W^{\alpha\beta\mu\nu} \\
{}^{(2)}W^{\alpha\beta\mu\nu} & = & \frac{1}{4} \epsilon^{\mu\nu}{}_{\rho\sigma} (\eta^{\alpha\rho} \Psi^{\beta\sigma} - \eta^{\beta\rho} \Psi^{\alpha\sigma}) \\
{}^{(3)}W^{\alpha\beta\mu\nu} & = & \frac{1}{12} X \epsilon^{\alpha\beta\mu\nu} \\
{}^{(4)}W^{\alpha\beta\mu\nu} & = & - \eta^{\mu[\alpha} \Phi^{\beta]\nu} + \eta^{\nu[\alpha} \Phi^{\beta]\mu} \\
{}^{(5)}W^{\alpha\beta\mu\nu} & = & \frac{1}{2} \left(\eta^{\alpha\mu} W_{\rm a}^{\beta\nu} - \eta^{\alpha\nu} W_{\rm a}^{\beta\mu} - \eta^{\beta\mu} W_{\rm a}^{\alpha\nu} + \eta^{\beta\nu} W_{\rm a}^{\alpha\mu}\right)  \\
{}^{(6)}W^{\alpha\beta\mu\nu} & = & - \frac{1}{6} W \eta^{\alpha [\mu} \eta^{|\beta|\nu]} \\
{}^{(1)}Z^{\alpha\beta\mu\nu} & = & Z^{\alpha\beta\mu\nu} - \sum_{i = 2}^5 {}^{(i)}Z^{\alpha\beta\mu\nu} \\
{}^{(2)}Z^{\alpha\beta\mu\nu} & = & \frac{1}{4} \epsilon^{\mu\nu}{}_{\rho\sigma} \left(\eta^{\alpha\rho} \Upsilon^{\beta\sigma} + \eta^{\beta\rho} \Upsilon^{\alpha\sigma}\right)  \\
{}^{(3)}Z^{\alpha\beta\mu\nu} & = &  \frac{1}{3} \left(\eta^{\alpha\mu} \Delta^{\beta\nu} - \eta^{\alpha\nu} \Delta^{\beta\mu} + \eta^{\beta\mu} \Delta^{\alpha\nu} - \eta^{\beta\nu} \Delta^{\alpha\mu} - \eta^{\alpha\beta} \Delta^{\mu\nu}\right) \\
{}^{(4)}Z^{\alpha\beta\mu\nu} & = & \frac{1}{4} \eta^{\alpha\beta} Z^{\mu\nu} \\
{}^{(5)}Z^{\alpha\beta\mu\nu} & = & \frac{1}{4} \left(\eta^{\alpha\mu} \Xi^{\beta\nu} - \eta^{\alpha\nu} \Xi^{\beta\mu} + \eta^{\beta\mu} \Xi^{\alpha\nu} - \eta^{\beta\nu} \Xi^{\alpha\mu}\right)
\end{eqnarray}
where we defined $W_{\rm a}^{\mu\nu} = W^{[\mu\nu]}$ as the
antisymmetric part of the Ricci tensor. From this the
decomposition (\ref{ConstGeneral}) follows.

\bibliography{qtn}

\end{document}